\documentclass[nofootinbib, twocolumn,superscriptaddress,preprintnumbers,letterpaper,nofootinbib,natbib]{revtex4-1}

\usepackage[english]{babel}

\usepackage{graphicx}
\usepackage{dcolumn}
\usepackage{bm}
\usepackage[breaklinks,colorlinks,urlcolor=blue,citecolor=blue,linkcolor=magenta]{hyperref}
\usepackage{epsfig}
\usepackage[usenames,dvipsnames]{xcolor}
\usepackage{multirow}
\usepackage{capt-of}
\usepackage{siunitx}
\usepackage{graphicx}
\usepackage{slashed}
\usepackage{tabularx}
\usepackage{braket}
\usepackage{amsfonts}
\usepackage{mathtools}
\usepackage{mathrsfs}
\usepackage{bbold}
\usepackage[compat=1.1.0]{tikz-feynman}
\usepackage{subfiles}
\usepackage{comment}

\usepackage{amsmath}
\usepackage{cleveref}

\usepackage{float}

\usepackage{slashed}
\usetikzlibrary{positioning}
\usepackage[compatibility=false]{caption}
\usepackage{subcaption}

\usepackage[normalem]{ulem}
\usepackage[shortlabels]{enumitem}  

\usepackage{fontawesome}
\setcitestyle{sort&compress}

\usepackage{float}

\pgfmathsetmacro\sizedot{1.1}
\pgfmathsetmacro\sizesqdot{1.5}
\pgfmathsetmacro\sizeemdot{2.}

\pgfmathsetmacro\sizecrodot{1.0}

{\left\lbrace\begin{array}{@{}l@{}}}%
{\end{array}\right.}
\usepackage{slashed}

\newcommand{\op}[2]{ \mathcal{O}_{#1} ^{#2} } 
\newcommand{\coeff}[2]{ c_{#1} ^{#2} } 
\newcommand{\RG}[1]{{\bf \color{red}{[RG: #1]}}}
\newcommand{\SDN}[1]{{\bf \color{violet}{[SDN: #1]}}}
\newcommand{\PR} {\mathbb{P}_R}
\newcommand{\PL} {\mathbb{P}_L}
\newcommand{\order}[1] {\mathcal{O}\left( #1 \right)}
\newcommand{\yuk}[1]{{Y}_{#1}}

\begin{document}
\preprint{KA-TP-18-2025}

\title{
Two loops, four tops and two $\gamma_5$ schemes: a renormalization story }
\author{Stefano Di Noi}
\email{stefano.dinoi@kit.edu} 
\affiliation{%
 Institute for Theoretical Physics, Karlsruhe Institute of Technology (KIT), D-76131 Karlsruhe, Germany
}
\author{Ramona Gr\"ober}%
 \email{ramona.groeber@pd.infn.it}
 \affiliation{%
 Dipartimento di Fisica e Astronomia ``G. Galilei", Universit\`a di Padova, Italy, \\ and Istituto
Nazionale di Fisica Nucleare, Sezione di Padova, I-35131 Padova, Italy
}
 
\date{\today}

\begin{abstract}
We calculate the four-top quark operator contributions to the two-loop renormalization constants of the fermion and gluon fields, necessary to obtain the renormalization group equations for the fermion masses and the strong coupling constant. 
The computation has been carried out in the na\"ive
dimensional regularization scheme for the $\gamma_5$ matrix. We also discuss a strategy to translate the results into the Breitenlohner-Maison-t’Hooft-Veltman  scheme and present the corresponding beta functions in both schemes.
\end{abstract}

\maketitle

\section{\label{sec:intro}Introduction}
Given the absence of a signal of light new physics, the use of Effective Field Theories (EFTs) as a bottom-up approach to new physics has become rather common. The Standard Model Effective Field Theory (SMEFT) assumes that the Standard Model (SM) particle content transforms in the same manner under its symmetries, but is augmented by higher-dimensional operators with respect to the SM. A basis of non-redundant operators at dimension-six, being the most relevant for collider physics, has been presented for the first time in Ref.~\cite{dim6smeft}.

With the ever-increasing precision of the LHC physics program, higher-order corrections to EFT predictions have become increasingly relevant. In particular, they can significantly enhance the sensitivity to new physics in EFT analyses, enlarging the set of observables which are sensitive to a given class of operators. This is especially true for Wilson coefficients that are currently only weakly constrained by direct searches. In such cases, loop-level contributions can play a crucial role in setting bounds—for instance, in the case of operators that modify the Higgs potential \cite{Degrassi:2016wml,  Gorbahn:2016uoy, Bizon:2016wgr, Maltoni:2017ims, DiVita:2017eyz, Degrassi:2019yix, terHoeve:2025omu, Haisch:2025pql} or four fermion operators containing four heavy fields \cite{deBlas:2015aea, Gauld:2015lmb, Dawson:2022bxd, Silvestrini:2018dos,  Alasfar:2022zyr, Degrande:2024mbg, Heinrich:2023rsd, Haisch:2024wnw, DiNoi:2025uhu}. 

The ultraviolet (UV) divergences of SMEFT determine the renormalization group equations (RGEs) of the SMEFT coefficients, which allow to efficiently resum the large logarithms appearing when there is a large separation of energy scales. At one-loop order at dimension-six the RGEs were given in Refs.~\cite{rge1, rge2, rge3}, implemented in various publicly available computer codes for a numeric solution \cite{Lyonnet:2013dna, Celis:2017hod, Aebischer:2018bkb, Fuentes-Martin:2020zaz, DiNoi:2022ejg}. Their effect on various processes at the LHC has been studied in Refs.~\cite{Grazzini:2018eyk, DiNoi:2023onw, Maltoni:2024dpn, Heinrich:2024rtg,DiNoi:2024ajj,Battaglia:2021nys}  and can also be significant in the context of global fits \cite{Bartocci:2024fmm, terHoeve:2025gey, deBlas:2025xhe}.
The full form of the anomalous dimension matrix at two-loop order at dimension-six in the SMEFT is still unknown, but various pieces exist in literature \cite{Bern:2020ikv, Jenkins:2023rtg, DiNoi:2023ygk, Jenkins:2023bls, Fuentes-Martin:2023ljp, DiNoi:2024ajj, Born:2024mgz, Duhr:2025zqw, Haisch:2025lvd, Assi:2025fsm}. Those two-loop RGE running effects can have a sizeable phenomenological impact \cite{DiNoi:2024ajj}, in particular for operators that upon matching to a UV model arise first at one-loop order in a renormalizable and weakly interacting theory \cite{Arzt:1994gp,Buchalla:2022vjp}. 

This work aims to provide yet another ingredient necessary to achieve the full two-loop anomalous dimension, namely the four-top quark operator contributions to the renormalization constants of the gluon and top quark field as well as the fermion mass counterterm. From their knowledge, the running of the strong coupling constant and the top quark mass can be obtained in a background field gauge \cite{Abbott:1981ke, Abbott:1983zw, Rebhan:1984bg}. In the same spirit, the contributions to these quantities stemming from the chromomagnetic operator and the operator modifying the gluon self-interactions have been recently computed in Refs.~\cite{Naterop:2024cfx,Duhr:2025zqw}. 

We note that in principle all light or light-heavy four fermion operators can contribute to the quantities we consider. The four-heavy operators, as already mentioned, are though much less constrained than the four-fermion operators involving light field. In light of this, for phenomenological reasons we concentrate on the four-top operators for now.  We note though that bounds on the operators $\op{QtQb}{(1,8)}$ are still rather weak \cite{Alasfar:2022zyr}, but as the contribution of these operators are usually suppressed as $m_b/m_t$ with respect to the four-top operators, we leave this computation for the future. To summarize, we include in our computation all the operators which generate at least a four-top vertex, which we refer to in the following as four-top operators. Some of these operators involve the chiral isodoublet, generating interactions with bottom quarks as well, which we include for consistency.

The four-top operator contributions contain chiral structures, which require the choice of a continuation scheme of $\gamma_5$ when employing dimensional regularization. Indeed,
the choice of the continuation scheme for $\gamma_5$ influences the functional form of the two-loop RGEs including four-fermion operators \cite{Ciuchini:1993fk,Ciuchini:1993fk,DiNoi:2023ygk}. 

In this paper, we will employ na\"ive dimensional regularization (NDR) \cite{Bardeen:1972vi, CHANOWITZ1979225}.
The NDR scheme can become inconsistent in the presence of six or more $\gamma$ matrices, hence it is often used together with reading point prescriptions \cite{Korner:1991sx, Kreimer:1993bh, Chen:2023lus, Chen:2024hlv}.
However, in the present case the computation is well-defined and non-ambigous, due to the relatively small number of $\gamma$ matrices which are present in the Dirac traces. 

Conversely, the Breitenlohner-Maison-t'Hooft-Veltman (BMHV) scheme \cite{Breitenlohner:1977hr, THOOFT1972189} is the only scheme that has been shown to be algebraically unambigiously defined to all loop orders 
\cite{Speer:1974cz,Breitenlohner:1975hg, Breitenlohner:1976te,Aoyama:1980yw,Costa:1977pd}. 
This is why ideally results should be presented in this computationally consistent yet more cumbersome scheme. 
In the BMHV scheme a split between the $4$-dimensional and the $(D-4)$-dimensional part of the $D$-dimensional Dirac algebra is performed, highly increasing the complexity of intermediate expressions. Additionally, this split induces a fictitious breaking of chiral symmetries originating in the fermion kinetic term. The symmetry can be restored adding finite counterterms, following, e.g., the method of Ref.~\cite{OlgosoRuiz:2024dzq} as we do in this paper. 

We show how the results can be obtained in the BMHV scheme by relying on the results in the NDR scheme.   
This connection is possible using the translation between the schemes presented in Ref.~\cite{DiNoi:2025uan}. 

The difference of the two-loop RGEs between NDR and BMHV has been studied extensively in the context of flavour physics \cite{Buras:1989xd, Buras:1991jm,Buras:1992tc,Ciuchini:1993ks,Ciuchini:1993fk} and in the SMEFT for single \cite{DiNoi:2023ygk} and double \cite{Heinrich:2023rsd} Higgs production.
Additionally, some results in the BMHV scheme have been presented for the Low Energy EFT (LEFT) \cite{Naterop:2023dek,Naterop:2024cfx,Naterop:2025lzc}.

The paper is structured as follows: in Sec.~\ref{sec:comp} we present our computational set-up, in Sec.~\ref{sec:NDR} we show our results in the NDR scheme and in Sec.~\ref{sec:BMHV} we present them in the BMHV scheme. Finally, in Sec.~\ref{sec:conclusion} we present our conclusions.

\section{Computational set-up \label{sec:comp}}
We consider here four heavy fermion operators which generate four-top effective vertices, following their definition in the Warsaw basis \cite{dim6smeft}:
\begin{equation} \label{eq:Lag4t}
\begin{split}
\mathcal{L}_{\text{4t}} &= \frac{\coeff{QQ}{(1)} }{\Lambda^2}\left(\bar{Q}_L \gamma_\mu Q_L\right)  \left(\bar{Q}_L \gamma^\mu Q_L \right) \\
&+\frac{\coeff{QQ}{(3)}}{\Lambda^2} \left(\bar{Q}_L \tau^I \gamma_\mu Q_L \right) \left(\bar{Q}_L \tau^I \gamma^\mu Q_L \right)  \\
&+ \frac{\coeff{Qt}{(1)}}{\Lambda^2}  \left(\bar{Q}_L \gamma_\mu Q_L \right) \left(\bar{t}_R \gamma^\mu t_R \right) \\
&+\frac{\coeff{Qt}{(8)}}{\Lambda^2} \left(\bar{Q}_L T^A\gamma_\mu Q_L \right) \left(\bar{t}_R T^A \gamma^\mu t_R \right) \\
&+ \frac{\coeff{tt}{}}{\Lambda^2} \left(\bar{t}_R  \gamma_\mu t_R \right) \left(\bar{t}_R \gamma^\mu t_R \right)\,.
\end{split}
\end{equation}
The field $t_R$ ($b_R$) stands for the right-handed top (bottom) quark field, while $Q_L=(t_L,b_L)^T$ stands for the $\mathrm{SU(2)_{L}}$ doublet of the third quark generation. The $\mathrm{SU(3)_C}$ generators are denoted as $T^A$ while $\tau^I$ are the Pauli matrices. We assume all the Wilson coefficients to be real.

We introduce the renormalization constants
\begin{equation} \label{eq:Zs}
\begin{split}
G^{(0)}=\sqrt{Z_G} G,\;& t^{(0)}_{L/R} =\sqrt{Z_t^{L/R}} t_{L/R},   \\
 m_t^{(0)}=Z_{m_t} m_t,\;g_s^{(0)}  & =Z_{g_s} g_s, \; \coeff{i}{(0)}=Z_{ij}\coeff{j}{},
\end{split}
\end{equation}
where $G,t,m_t,g_s,\coeff{i}{}$ are the renormalized quantities while their counterparts with the superscript $(0)$ represent the bare quantities. 
The renormalization constants with a single insertion of a four-top operator can be written as a series expansion in the coupling constants up to dimension six. In this work, we will adopt, schematically, the following expansion
\begin{equation}
Z_{\mathrm{X}}=1+\delta \mathrm{X}, \quad \delta \mathrm{X} =\order{\dfrac{g_s^2}{(16\pi^2)^2} \times \frac{1}{\Lambda^2}},
\end{equation}
accounting only for QCD corrections. 

Since we are working in the broken phase, we have \cite{Jenkins:2017jig}:
\begin{equation} \label{eq:defgs}
g_s = g_3 \left(1 + g_3^2 \frac{v^2}{\Lambda^2} \coeff{HG}{} \right).
\end{equation}
In the previous expression, $g_3$ refers to the pure $\mathrm{SU(3)}$ gauge coupling, while $g_s$ is the effective gauge coupling below the EW scale. 

We generate the Feynman diagrams with \texttt{qgraf-3.6.10} \cite{Nogueira:1991ex}, perform the Dirac algebra with \texttt{FeynCalc} \cite{Mertig:1990an, Shtabovenko:2016sxi, Shtabovenko:2020gxv}  and reduce to Master Integrals (MIs) via Integration By Parts (IBP) identities with \texttt{LiteRed} \cite{Lee:2012cn, Lee:2013mka} using the same tool chain as in Ref.~\cite{DiNoi:2023ygk}.
The thus obtained MIs can be found in Ref.~\cite{Martin:2005qm}. We employ dimensional regularization with $D=4-2\epsilon$ in the $\overline{\mathrm{MS}}$-scheme.

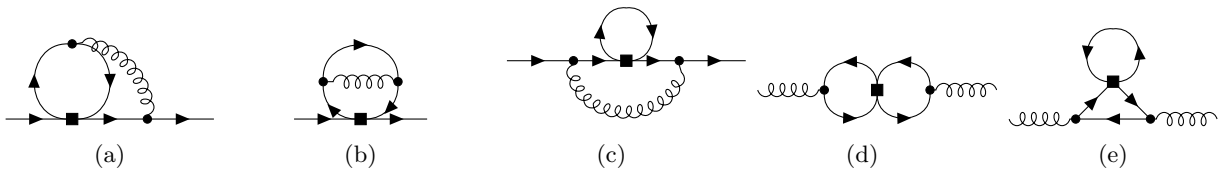
\begin{figure*}[!ht]
\centering
    \begin{subfigure}{0.18\textwidth}
        \centering
       \begin{tikzpicture} 
            \begin{feynman}[small]
                \vertex  (t1)  {};
                \vertex  (4t) [square dot,scale=\sizesqdot,right = of t1] {};
                \vertex (g1) [dot, scale=\sizedot,above= of 4t,color=black] {};
                \vertex  (g2) [dot,scale=\sizedot,right = of 4t] {};
                \vertex  (t2) [right=of g2]  {};
                \diagram* {
                    (t1)  -- [fermion] (4t) -- [fermion] (g2) -- [fermion] (t2),
                    (g2) -- [gluon, quarter right] (g1),
                   (4t) -- [fermion, half left] (g1) -- [fermion, half left] (4t),
                };
            \end{feynman}
        \end{tikzpicture}
        \caption{} \label{fig:props-t-peng}
    \end{subfigure}
\begin{subfigure}{0.18\textwidth}
        \centering
       \begin{tikzpicture} 
            \begin{feynman}[small]
                \vertex  (t1)  {};
                \vertex  (4t) [square dot,scale=\sizesqdot,right = of t1] {};
                \vertex (g1) [dot, scale=\sizedot,above left = 20 pt of 4t,color=black] {};
                \vertex  (g2) [dot,scale=\sizedot, above right = 20 pt of 4t] {};
                \vertex  (g3) [dot,scale=0.001, above left =20 pt of g2] {};
                \vertex  (t2) [right=of 4t]  {};
                \diagram* {
                    (t1)  -- [fermion] (4t)  -- [fermion] (t2),
                    (g2) -- [gluon] (g1),
                   (4t) -- [fermion, quarter left] (g1) -- [fermion, half left] (g2) -- [fermion, quarter left] (4t),
                };
            \end{feynman}
        \end{tikzpicture}
        \caption{}\label{fig:props-t-4t}

    \end{subfigure}
   \begin{subfigure}{0.18\textwidth}
        \centering
       \begin{tikzpicture}[baseline=(g3)]
            \begin{feynman}[small]
                \vertex  (t1)  {};
                \vertex (g1) [dot, scale=\sizedot,right = of t1,color=black] {};
                 \vertex  (4t) [square dot,scale=\sizesqdot,right = 20 pt of g1] {};
                \vertex  (g2) [dot,scale=\sizedot, right = 20 pt  of 4t] {};
                \vertex  (g3) [dot,scale=0.01, above = 20 pt of 4t] {};
                \vertex  (t2) [right=of g2]  {};
                \diagram* {
                    (g1)  -- [gluon, half right] (g2),
                    (t1) -- [fermion] (g1)-- [fermion] (4t)-- [fermion] (g2)-- [fermion] (t2),
                   (4t) -- [fermion, half left] (g3) -- [fermion, half left] (4t),
                   };
            \end{feynman}
        \end{tikzpicture}
        \caption{} \label{fig:props-t-Z2}

    \end{subfigure}
     \begin{subfigure}{0.18\textwidth}
        \centering
       \begin{tikzpicture} 
            \begin{feynman}[small]
                \vertex  (t1)  {};
                \vertex (g1) [dot, scale=\sizedot,right = of t1,color=black] {};
                 \vertex  (4t) [square dot,scale=\sizesqdot,right = 20 pt of g1] {};
                \vertex  (g2) [dot,scale=\sizedot, right = 20 pt  of 4t] {};
                \vertex  (t2) [right=of g2]  {};
                \diagram* {
                    (t1)  -- [gluon] (g1),(t2)  -- [gluon] (g2),
                    (g1) -- [fermion,half right] (4t)-- [fermion, half right] (g1),
                    (g2) -- [fermion,half right] (4t)-- [fermion, half right] (g2),
                   };
            \end{feynman}
        \end{tikzpicture}
        \caption{} 

    \end{subfigure}
     \begin{subfigure}{0.18\textwidth}
        \centering
     \begin{tikzpicture} 
            \begin{feynman}[small]
                \vertex  (t1)  {};
                \vertex (g1) [dot, scale=\sizedot,right = of t1,color=black] {};
                 \vertex  (4t) [square dot,scale=\sizesqdot,above right = 20 pt of g1] {};
                \vertex  (g2) [dot,scale=\sizedot,below  right = 20 pt of 4t] {};
                \vertex  (g3) [dot,scale=0.01, above = 20 pt  of 4t] {};
                \vertex  (t2) [right=of g2]  {};
                \diagram* {
                    (t1)  -- [gluon] (g1),(t2)  -- [gluon] (g2),
                    (g3) -- [fermion,half right] (4t)-- [fermion, half right] (g3),
                    (g1) --[fermion] (4t) --[fermion] (g2) --[fermion] (g1),
                   };
            \end{feynman}
        \end{tikzpicture}
        \caption{} 

    \end{subfigure}
\caption{Representative diagrams displaying the four-top operators (solid square dot) to the top quark propagator (a-c) and to the gluon propagator (d-e). The internal fermion line can represent either a top or a bottom quark.}  \label{fig:props}
\end{figure*}

We show a sample of the two-loop diagrams of order $\mathcal{O}(\alpha_s \coeff{4t}{})$ for the gluon vacuum polarisation and for the self-energy of the top quark in Fig.~\ref{fig:props}.

We perform the renormalization in the off-shell basis, meaning that the kinematic-dependent divergences must be removed by operators which are then removed by the Equation of Motion (EOMs). 

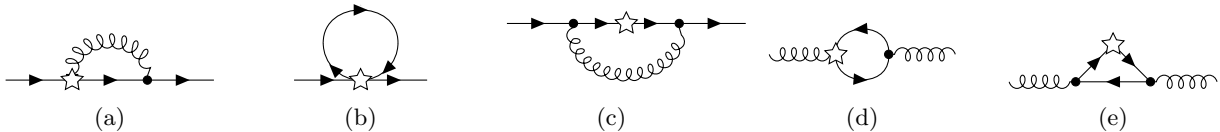
\begin{figure*}[!ht]
\centering
    \begin{subfigure}{0.18\textwidth}
        \centering
       \begin{tikzpicture} 
            \begin{feynman}[small]
                \vertex  (t1)  {};
                \vertex  (4t) [square dot,scale=\sizesqdot,right = of t1] {};
                \vertex  (g2) [dot,scale=\sizedot,right = of 4t] {};
                \vertex  (t2) [right=of g2]  {};
                \diagram* {
                    (t1)  -- [fermion] (4t) -- [fermion] (g2) -- [fermion] (t2),
                    (4t) -- [gluon, half left] (g2),
                };
                 \node[shape=star,star points=5,star point ratio = 2,fill=white, draw,scale = 0.5,opacity=1] at (4t) {};
            \end{feynman}
        \end{tikzpicture}
        \caption{}\label{fig:propsct-t-peng}
    \end{subfigure}
\begin{subfigure}{0.18\textwidth}
        \centering
       \begin{tikzpicture} 
            \begin{feynman}[small]
                \vertex  (t1)  {};
                \vertex  (4t) [square dot,scale=0.01,right = of t1] {};
                \vertex (g1) [dot, scale=0.01,above left = 20 pt of 4t,color=black] {};
                \vertex  (g2) [dot,scale=0.01, above right = 20 pt of 4t] {};
                \vertex  (g3) [dot,scale=0.001, above left =20 pt of g2] {};
                \vertex  (t2) [right=of 4t]  {};
                \diagram* {
                    (t1)  -- [fermion] (4t)  -- [fermion] (t2),
                   (4t) -- [fermion, quarter left] (g1) -- [fermion, half left] (g2) -- [fermion, quarter left] (4t),
                };
                 \node[shape=star,star points=5,star point ratio = 2,fill=white, draw,scale = 0.5,opacity=1] at (4t) {};
            \end{feynman}
        \end{tikzpicture}
        \caption{} \label{fig:propsct-t-4t}

    \end{subfigure}
   \begin{subfigure}{0.18\textwidth}
        \centering
       \begin{tikzpicture}[baseline=(g3)]
            \begin{feynman}[small]
                \vertex  (t1)  {};
                \vertex (g1) [dot, scale=\sizedot,right = of t1,color=black] {};
                 \vertex  (4t) [square dot,scale=0.1,right = 20 pt of g1] {};
                \vertex  (g2) [dot,scale=\sizedot, right = 20 pt  of 4t] {};
                \vertex  (g3) [dot,scale=0.01, above = 20 pt of 4t] {};
                \vertex  (t2) [right=of g2]  {};
                \diagram* {
                    (g1)  -- [gluon, half right] (g2),
                    (t1) -- [fermion] (g1)-- [fermion] (4t)-- [fermion] (g2)-- [fermion] (t2),
                   };
            \end{feynman}
             \node[shape=star,star points=5,star point ratio = 2,fill=white, draw,scale = 0.5,opacity=1] at (4t) {};
        \end{tikzpicture}
        \caption{} \label{fig:propsct-t-Z2}

    \end{subfigure}
     \begin{subfigure}{0.18\textwidth}
        \centering
       \begin{tikzpicture} 
            \begin{feynman}[small]
                \vertex  (t1)  {};
                 \vertex  (4t) [square dot,scale=0.01,right = of t1] {};
                \vertex  (g2) [dot,scale=\sizedot, right = 20 pt  of 4t] {};
                \vertex  (t2) [right=of g2]  {};
                \diagram* {
                    (t1)  -- [gluon] (g1),(t2)  -- [gluon] (g2),
                    (g2) -- [fermion,half right] (4t)-- [fermion, half right] (g2),
                   };
            \end{feynman}
             \node[shape=star,star points=5,star point ratio = 2,fill=white, draw,scale = 0.5,opacity=1] at (4t) {};
        \end{tikzpicture}
        \caption{} \label{fig:propsct-g-peng}

    \end{subfigure}
     \begin{subfigure}{0.18\textwidth}
        \centering
     \begin{tikzpicture} 
            \begin{feynman}[small]
                \vertex  (t1)  {};
                \vertex (g1) [dot, scale=\sizedot,right = of t1,color=black] {};
                 \vertex  (4t) [square dot,scale=0.01,above right = 20 pt of g1] {};
                \vertex  (g2) [dot,scale=\sizedot,below  right = 20 pt of 4t] {};
                \vertex  (g3) [dot,scale=0.01, above = 20 pt  of 4t] {};
                \vertex  (t2) [right=of g2]  {};
                \diagram* {
                    (t1)  -- [gluon] (g1),(t2)  -- [gluon] (g2),
                    (g1) --[fermion] (4t) --[fermion] (g2) --[fermion] (g1),
                   };
            \end{feynman}
             \node[shape=star,star points=5,star point ratio = 2,fill=white, draw,scale = 0.5,opacity=1] at (4t) {};
        \end{tikzpicture}
        \caption{} \label{fig:propsct-g-Z2}

    \end{subfigure}
\caption{Representative diagrams displaying the one-loop insertion of the one-loop counterterms to the top quark propagator (a-c) and to the gluon propagator (d-e). } \label{fig:propsct}
\end{figure*}

We show in Fig.~\ref{fig:propsct} a sample of the one-loop insertions of one-loop counterterms. In particular, in Figs.~\ref{fig:propsct-t-peng} and \ref{fig:propsct-g-peng} we show an insertion of the counterterm of the operators $\mathcal{R}_{GQ}$,~$\mathcal{R}_{Gb}$,~$\mathcal{R}_{Gt}$ \cite{Carmona:2021xtq}
\begin{align}
\mathcal{R}_{GQ}&=(\bar{Q}_LT^A\gamma^{\mu}Q_L)D^{\nu}G^{A}_{\mu\nu}, \label{eq:RGQ} \\ 
\mathcal{R}_{Gb}&=(\bar{b}_RT^A\gamma^{\mu}b_R)D^{\nu}G^{A}_{\mu\nu}\,\label{eq:RGb} ,\\
\mathcal{R}_{Gt}&=(\bar{t}_RT^A\gamma^{\mu}t_R)D^{\nu}G^{A}_{\mu\nu}\,\label{eq:RGt},
\end{align}
 which generate an off-shell gluon vertex \cite{Alasfar:2022zyr}:\footnote{In this reference, the contribution from the top-loop only is given.} \\
\begin{minipage}{.07 \textwidth}
\begin{tikzpicture}[baseline=(4t)] 
            \begin{feynman}[small]
                \vertex  (t1)  {};
                \vertex  (4t) [square dot,scale=0.01,right = of t1] {};
                \vertex  (g2) [above = of 4t] {$g$};
                \vertex  (t2) [right=of 4t]  {};
                \diagram* {
                    (t1)  -- [fermion] (4t)  -- [fermion] (t2),
                    (g2) -- [gluon] (4t),
                };
            \end{feynman}
             \node[shape=star,star points=5,star point ratio = 2,fill=white, draw,scale = 0.5,opacity=1] at (4t) {};
        \end{tikzpicture}
\end{minipage}
\begin{minipage}{.4 \textwidth}
\begin{equation} 
\begin{aligned} \label{eq:ctPenguin}
&\quad -i\frac{g_s^2}{48 \pi^2 \epsilon} \left(p^2 g^{\mu \nu} - p^\mu p^\nu \right) \gamma_\nu   \\ 
=& \quad \times \Big[  \left(4 \coeff{QQ}{(1)}+12 \coeff{QQ}{(1)} + \coeff{Qt}{(8)} \right) \PL  \\ 
&\quad + \; \; \,\left(4 \coeff{tt}{} +  2\coeff{Qt}{(8)} \right) \PR
        \Big].
\end{aligned}
\end{equation}
\end{minipage}

In Fig.~\ref{fig:propsct-t-4t} we show an insertion of the counterterm associated to the four-top operators. The counterterm can be obtained by the $\order{g_s^2}$ anomalous dimension matrix of the SMEFT \cite{rge3}, using 
\begin{equation}\label{eq:ctfromRGE}
\delta \coeff{i}{}  = \frac{1}{{\epsilon}} \frac{1}{2 L }\gamma_{ij}\coeff{j}{} ,
\end{equation}
being $L$ the number of loops and $\beta_i\equiv \gamma_{ij} \coeff{j}{} \equiv \mu \dfrac{d \coeff{i}{}}{d\mu}$ with $i,j=\text{}^{(1),(8)}_{Qt}, \text{}_{QQ}^{(1),(3)}, tt$.
We highlight that the counterterms we consider do not include the indirect EOM-induced contributions. Such terms arise when the operators in Eqs.~(\ref{eq:RGQ}-\ref{eq:RGt}) are removed using the SM EOMs in favour of four-top operators \cite{Silvestrini:2018dos}, which is not the case in our off-shell computation. 
The indirect contributions are easily
identified due to their peculiar penguin-like flavour structure with a sum over two flavour indices. We stress that these terms should not be included here since the penguin-like subdivergences are renormalized with the off-shell counterterm in Eq.~\eqref{eq:ctPenguin}. 

Additionally, the RGE contains also the contribution stemming from the wavefunction renormalization constant of the external fields, which must be considered. 
In other words, the four-top counterterm should include only the contributions of the one-particle irreducible $\order{\alpha_s}$ diagrams, which we have checked with an explicit computation. 

Finally, we report in Fig.~\ref{fig:propsct-t-Z2},~\ref{fig:propsct-g-Z2} an insertion of $\delta m_t$ proportional to the four-top operators, given by \cite{DiNoi:2023ygk}\footnote{The sign difference arises from the different convention chosen for the mass counterterm $\delta m_t$.}
\begin{equation}
\delta m_t = - \frac{m_t^2}{4 \pi^2 \Lambda^2 \epsilon} \left( \coeff{Qt}{(1)}+\frac{4}{3}\coeff{Qt}{(1)} \right).
\end{equation}

The combination of the two-loop bare amplitudes with the one-loop insertions of the one-loop counterterms yields a local\footnote{ The locality for the counterterm can be confirmed by checking that the double pole arising from the two-loop diagrams (see Fig.~\ref{fig:props}) is -1/2 of the double pole of the diagrams arising from the one-loop insertion of the one-loop counterterms (see Fig.~\ref{fig:propsct}), as we did in our computation. }  but kinematics-dependent divergent part, which cannot be reabsorbed by the SM renormalization constants alone. 
In order to renormalize such contribution, redundant operators are needed. In this work we employ the following operators: 
\begin{align}
\mathcal{R}_{\partial ^2 G} &=\Big(D^\mu G^A_{\mu \nu} 
\Big)^2 \, ,\label{eq:Rd2G} \\ 
\mathcal{R}_{\partial ^2 t} &= m_t \, \bar{t} \left(i \slashed{D} -m \right)^2 t  \, \label{eq:Rd2t} ,\\ 
\mathcal{R}_{\partial ^3 t R} &= \bar{t}_R (\partial^\mu \partial_\mu + m^2) i \slashed{D} t_R \,\label{eq:Rd3tR} , \\ 
\mathcal{R}_{\partial ^3 t L} &= \bar{t}_L (\partial^\mu \partial_\mu + m^2) i \slashed{D}  t_L \,\label{eq:Rd3tL} .
\end{align}
We notice that, for the purpose of the gluon propagator renormalization, the choice $\mathcal{R}_{\partial ^2 G} = \left( D^\mu G^A_{\mu \nu} 
- g_s {\textstyle\sum_\psi} \bar{\psi} \gamma_\nu T^A \psi
\right)^2$ of Ref.~\cite{Duhr:2025zqw} (which we follow for the definition of $\mathcal{R}_{\partial^2 t}$) would be equivalent. However, this choice would then feed back in Eqs.~(\ref{eq:RGQ}-\ref{eq:RGt}). 
We stress that the choice of redundant operators is non-unique, affecting the final results for the RGEs. For this reason, we present in App.~\ref{app:Amplitude} the two-point functions where no renormalization of redundant operators at two-loop level has been performed, allowing for a different basis choice and future comparison. Additionally, we present the results in the on-shell basis, where no redundant operators are present, in App.~\ref{app:OSBasis}. 
We stress that employing an on-shell EFT basis allows to use a minimal and non-redundant set of operators. Hence after reducing to such a basis, the differences in the choice of redundant operators has to vanish.

In principle, also the contribution from the operator $\mathcal{R}_{Qt}^{(1)}=(\bar{Q}_L t_R)(\bar{t}_R Q_L)$ should be considered. This operator can be reduced to the Warsaw basis by means of the one-loop corrected Fierz identities \cite{Fuentes-Martin:2022vvu}, projecting into the Yukawa operator and to $\op{tH}{}\equiv \op{uH}{33}$, which are connected to the top mass via 
\begin{equation} \label{eq:mtSMEFT}
m = \frac{v}{\sqrt{2}} \left( \yuk{t} -\frac{v^2}{2} \frac{\coeff{t H}{}}{\Lambda^2} \right).
\end{equation}

However, $\mathcal{R}_{Qt}^{(1)}$ is not generated by the renormalization of four-top operators at one-loop and thus this contribution does not need to be included. 

The wave function renormalization constants of the gluon and can be obtained from the transversal part $\Pi_{\mathrm{T}}(p^2)$ of the vacuum polarisation, defined as 

\begin{equation}
\Pi_{\mu \nu}(p^2) = \Pi_{\mathrm{T}}(p^2) \left(  g_{\mu \nu} -\frac{ p_\mu p_\nu}{p^2} \right). 
\end{equation}
The renormalization constant in the $\overline{\mathrm{MS}}$-scheme is defined as 
\begin{equation}
\delta Z_G=-\frac{\partial \Pi_{\mathrm{T}}(p^2)}{\partial p^2}\Bigg|^{\text{div}}_{p^2=0}\,.
\end{equation}
We notice that setting $p^2 = 0$ separates the contribution to $\delta Z_G$ from the renormalization of $\mathcal{R}_{\partial ^2 G}$. The latter quantity would give a universal contribution to the two-loop running of four-quark operators at $\order{g_s^4}$ with the schematic structure $\left(\bar{\chi}T^A \gamma^\mu \chi \right)\left(\bar{\psi}T^A \gamma_\mu \psi \right)$. However, such computation is beyond the scope of this paper. 

The fermion propagator can be decomposed in three scalar form-factors as 
\begin{equation} \label{eq:Sigma}
\Sigma(p^2)=  \slashed{p} \PL \Sigma_L(p^2)+\slashed{p} \PR \Sigma_R(p^2) +  m \Sigma_S(p^2) .
\end{equation}

In the fermion sector the separation between the contributions to the wavefunction renormalization constant and the EOM-vanishing redundant operators in Eqs.~(\ref{eq:Rd2t}-\ref{eq:Rd3tL}) is less straightforward than for the gluon. The reason is that the redundant operators not only contribute   with terms which are $\propto p^2$ but also with $m^2$. We present the results in the NDR scheme for the counterterms of the operators in Eqs.~(\ref{eq:Rd2t}-\ref{eq:Rd3tL}) in the following section.

\section{Results in the NDR scheme \label{sec:NDR}}
In first instance, we computed our result in the NDR scheme for $\gamma_5$ which, in this case, does not give rise to ambiguities. The contributions from four-top operators to the gluon propagator can generate either one or two traces in Dirac space. In the former case, we can ignore the trace with a single $\gamma_5$ (which may be ambigous) since it does not contribute to $Z_G$.\footnote{Such a contribution would renormalize the operator $\op{H\tilde{G}}{}$, which we do not consider here.} Conversely, the contributions with two traces never reach the number of 6 $\gamma$ matrices and a $\gamma_5$, hence do not give rise to any ambiguities. 

For what concerns the top quark propagator, there can be either one or zero traces. In the former case we can ignore the traces with a single $\gamma_5$ as they do not contribute. In particular, the diagram in Fig.~\ref{fig:props-t-4t} cannot generate terms proportional to the Levi-Civita pseudotensor by Lorentz symmetry, while the diagram in Fig.~\ref{fig:props-t-Z2} has at most 3 $\gamma$ matrices and a $\gamma_5$.
The penguin-like diagram in Fig.~\ref{fig:props-t-peng} could contribute with terms with the structure $\epsilon^{\mu \nu \alpha \beta} \gamma_\mu p_\nu l_\alpha k_\beta$, with $l,k$ being the loop momenta, but cancel when the the diagram in Fig.~\ref{fig:props-t-peng} is summed with the corresponding diagram where the gluon propagator is connected on the left.\footnote{An alternative argument is that they should also give zero after integration by parts identities and loop integration is applied as there are not sufficient momenta to build a completely antisymmetric object multiplying the $\epsilon^{\mu \nu \alpha \beta}$.} 
Finally, the case with no traces is trivially non-ambiguous.

A clear advantage of the NDR scheme is that much less cumbersome Dirac algebra has to be applied, allowing also to directly reduce the obtained expressions to master intergrals via the usual tool chain.\footnote{Due to the splitting in 4 and $\epsilon$ dimensions in the BMHV scheme, additional steps need to be undergone there \cite{Heller:2020owb,vonManteuffel:2025swv}.} 

We present the counterterms of the operators defined in Eqs.~(\ref{eq:Rd2t}-\ref{eq:Rd3tL}), defining $ L_2 \equiv  { g_s^2}/{(16 \pi^2)^2}$ for brevity:
\begin{equation}
\delta \mathcal{R}_{\partial ^2 t} =\frac{1}{3 \epsilon} L_2 \Big[
4\coeff{QQ}{(1)}+12 \coeff{QQ}{(3)}+4 \coeff{tt}{} + 12 \coeff{Qt}{(1)} + \coeff{Qt}{(8)}
\Big]  \, \label{eq:ctRd2t}, 
\end{equation}
\begin{equation}
\delta \mathcal{R}_{\partial ^3 t R} = L_2   \Big[\frac{4}{9 \epsilon^2}(\coeff{Qt}{(8)}+2\coeff{tt}{})  + \frac{2}{9 \epsilon}(2\coeff{tt}{}-\coeff{Qt}{(8)}) \Big]  \,\label{eq:ctRd3tR},
\end{equation}
\begin{equation}
\begin{aligned}
\delta \mathcal{R}_{\partial ^3 t L } = L_2 &\Big[\frac{2}{9 \epsilon^2}(\coeff{Qt}{(8)}+4\coeff{QQ}{(1)}+12\coeff{QQ}{(3)})  \\ &+\frac{1}{9 \epsilon}(4\coeff{QQ}{(1)}+12\coeff{QQ}{(3)}-\coeff{Qt}{(8)})   \Big]  \,\label{eq:ctRd3tL}.
\end{aligned}
\end{equation}

We find for the wave function renormalization constants of the top quark field
\begin{align}
 \delta Z_t^{R} = \frac{L_2 m_t^2}{\Lambda^2} &\Bigg[\frac{8}{9\epsilon^2}(\coeff{Qt}{(8)}+2\coeff{tt}{}) \\
 &+ \frac{4}{9\epsilon}(9\coeff{Qt}{(1)}+4\coeff{Qt}{(8)} \\ &+12\coeff{tt}{}+6\coeff{QQ}{(1)}+18\coeff{QQ}{(3)}) \Bigg],\label{eq:d2R}\\ 
 \delta Z_t^{L} = \frac{L_2 m_t^2}{\Lambda^2} &\Bigg[\frac{4}{9\epsilon^2}(\coeff{Qt}{(8)}+4\coeff{QQ}{(1)}+12 \coeff{QQ}{(3)}) \\
 &+ \frac{2}{9\epsilon}(18\coeff{Qt}{(1)}+7\coeff{Qt}{(8)} \\ &+12\coeff{tt}{}+24\coeff{QQ}{(1)}+72\coeff{QQ}{(3)}) \Bigg].\label{eq:d2L} 
\end{align}

Finally, for the fermion mass counterterm we find
\begin{equation}\label{eq:dm}
\begin{aligned}
 \delta m_t = \frac{L_2 m_t^2}{\Lambda^2} \Bigg[&\frac{8}{9\epsilon^2}(
 60\coeff{Qt}{(1)}+77\coeff{Qt}{(8)}
- 4 \coeff{QQ}{(1)}-12\coeff{QQ}{(3)} -4 \coeff{tt}{}
 ) \\
 &+ \frac{m_b^2}{m_t^2 }\frac{4}{\epsilon}(8 \coeff{QQ}{(3)}+\coeff{Qt}{(8)})\\
 &+ \frac{8}{9 \epsilon}(16\coeff{QQ}{(1)}+12\coeff{QQ}{(3)}+16 \coeff{tt}{} \\ &-3\coeff{Qt}{(1)} -34 \coeff{Qt}{(8)})
 \Bigg].
\end{aligned}
\end{equation}

The wave function renormalization constant for the gluon field is
\begin{equation}\label{eq:ZG}
\delta Z_G =  \frac{4 g_s^2 m^2}{(16 \pi^2)^2 \Lambda^2} \frac{1}{\epsilon} \left( \coeff{Qt}{(1)} - \frac{1}{6} \coeff{Qt}{(8)} \right)\,.
\end{equation}

By using $Z_G \sqrt{Z_{g_s}} = 1$ holding in the background-field gauge \cite{Bohm:2001yx}, we have
\begin{equation}\label{eq:Zgs}
\delta {g_s} =  -\frac{2 g_s^2 m^2}{(16 \pi^2)^2 \Lambda^2} \frac{1}{\epsilon} \left( \coeff{Qt}{(1)} - \frac{1}{6} \coeff{Qt}{(8)} \right).
\end{equation}
 While we did not explicitly use background field gauge, in our case given that we have no diagrams with ghost fields at the considered order and only external gluon fields in the gluon propagator, it effectively does not make any difference.

We can finally extract the RGE from the coefficient of the single pole of the counterterm, multiplying it by $2L$,  being $L$ the number of loops.
By defining
\begin{equation}
\beta_X \equiv \mu \dfrac{dX}{d \mu}    
\end{equation}
we find
\begin{equation} \label{eq:betamtNDR}
\begin{aligned}
\beta_{m_t}^{\mathrm{NDR}} &=L_2 \frac{m_t^3}{\Lambda^2} \frac{32}{9} (16 \coeff{tt}{}+16 \coeff{QQ}{(1)} + 12 \coeff{QQ}{(3)} -3 \coeff{Qt}{(1)}-34 \coeff{Qt}{(8)}) \\ 
&+ L_2 \frac{m_t m_b^2}{\Lambda^2} 16 (8  \coeff{QQ}{(3)}+ \coeff{Qt}{(8)}).
\end{aligned}
\end{equation}

Regarding the running of the strong coupling, we have: 
\begin{equation}\label{eq:betagsNDR}
\beta_{g_s}^{\mathrm{NDR}} =  -\frac{8 g_s^2 m_t^2}{(16 \pi^2)^2 \Lambda^2}\times \left( \coeff{Qt}{(1)} - \frac{1}{6} \coeff{Qt}{(8)} \right).
\end{equation}
We observe that this result agrees with the one that can be obtained from Eq.~\eqref{eq:defgs}, which yields 
\begin{equation}
\beta_{g_s} = \beta_{g_3} + g_s \dfrac{v^2}{\Lambda^2} \beta \coeff{HG}{}.
\end{equation} 
The contribution from four-top operators to $g_3$ in the unbroken phase
is proportional to $s^2 (g^{\mu\nu} - q^\mu q^\nu/s)$, following from dimensional analysis.   
These contributions renormalize the redundant operator $\mathcal{R}_{\partial^2 G}$ but not $Z_G$, from which it follows that the contribution to the first term vanishes. 

We can extract the 
final result from 
\begin{equation}
\beta \coeff{HG}{}=- \dfrac{8  m^2 g_s}{(16 \pi^2)^2  v^2}  \left( \coeff{Qt}{(1)} - \dfrac{1}{6} \coeff{Qt}{(8)} \right),
\end{equation}
first provided in Ref.~\cite{DiNoi:2023ygk}, confirming the result in Eq.~\eqref{eq:betagsNDR}.

\section{Results in the BMHV scheme \label{sec:BMHV}}
A direct computation in the BMHV scheme might necessitate the insertion of a large amount of evanescent operators given that all Lorentz indices should be split into 4 and $2\epsilon$ dimensional pieces \cite{Naterop:2023dek}.  In addition, in the BMHV scheme the regulator breaks chiral symmetries that can be restored  by opportune finite counterterms, see for instance Ref.~\cite{OlgosoRuiz:2024dzq} for a method.   
We avoid such difficulties by using the translation table between the BMHV  and NDR scheme given in Ref.~\cite{DiNoi:2025uan}. 
This reference presents the one-loop shifts in the Wilson coefficients between BMHV and NDR, making it possible to obtain the RGEs in the BMHV scheme by combining the NDR result and the one-loop shifts inserted in the one-loop RGEs \cite{rge1,rge2,rge3}. We emphasise that the one-loop translations are sufficient for our purpose even though we work at two-loop order: the two-loop shifts would only be needed for when a pole in $1/\epsilon^2$ meets the $\order{\epsilon}$ term in the RGE. But since those poles at two-loop order do not depend on the continuation scheme, the one-loop translation table is sufficient. We follow here the symmetry-restoring approach, even though it is not mandatory.\footnote{Since we work in the broken phase the chiral symmetry is anyways broken by the fermion masses.}

We would like to note that \textit{in general} this procedure cannot be applied, due to inconsistencies arising in the NDR scheme. However, the validity of the procedure in the present case is assured by the small amount of $\gamma $ matrices involved in our calculation, as discussed in the previous section. A translation procedure between the two schemes was similarly applied in the context of the LEFT in Ref.~\cite{Naterop:2025lzc}. 

For the running of the strong coupling, the only shift proportional to the four-top operator that enters our computation is the one affecting the chromomagnetic operator $\op{tG}{}\equiv \op{uG}{33}$. Using the shift together with Eq.~\eqref{eq:betagsNDR}, we find 
\begin{equation}\label{eq:betagsBMHV}
\beta_{g_s}^{\mathrm{BMHV}} =  0.
\end{equation}
This result agrees once again with the findings of Ref.~\cite{DiNoi:2023ygk}, where it is shown that four-top quark operators do not contribute to the running of $\op{HG}{}$ at two-loop level.\footnote{We thank the authors of Refs.~\cite{Venturainprep, Nateropinprep} for confirming this result.}

For what concerns the running of the top mass, we consider the shifts affecting all the operators entering the RGEs of the Yukawa coupling \cite{rge1} and $\coeff{tH}{}\equiv \coeff{uH}{3,3}$ \cite{rge1,rge2,rge3}, which are related to the mass as prescribed by Eq.~\eqref{eq:mtSMEFT}. However, the shift to the Yukawa coupling do not contribute at the present order, being $\order{ \lambda }$.

For our computation at $\order{g_s^2}$, we need the shifts for the operators $\op{Qt}{(1,8)}$,~$\op{tH}{}$ and $\op{tG}{}\equiv \op{uH}{33}$, leading to 
\begin{equation}\label{eq:betamtshift}
\begin{split}
\Delta \beta_{m_t} = -\frac{16}{9}L_2 \frac{m_t^3}{\Lambda^2} & \Big(8 \coeff{tt}{}+20 \coeff{QQ}{(1)}+24  \coeff{QQ}{(3)} \\ &-24  \coeff{Qt}{(1)}-23 \coeff{Qt}{(8)}\Big).
\end{split}
\end{equation}
We finally obtain, by using $\beta_{m_t}^{\mathrm{BMHV}} = \beta_{m_t}^{\mathrm{NDR}}+\Delta \beta_{m_t}$: 
\begin{equation} \label{eq:betamtBMHV}
\begin{aligned}
\beta_{m_t}^{\mathrm{BMHV}} &= 
 L_2 \frac{m_t^3}{\Lambda^2} \frac{16}{3} (4 \coeff{QQ}{(1)} + 8 \coeff{tt}{} + 6 \coeff{Qt}{(1)} - 15 \coeff{Qt}{(8)}) \\ 
 &+ L_2 \frac{m_t m_b^2}{\Lambda^2} 16 (8 \coeff{QQ}{(3)}+\coeff{Qt}{(8)})
.     
\end{aligned}
\end{equation}
The differences between BMHV and NDR RGEs can cancel upon matching to a UV theory due to scheme-dependent matching coefficients \cite{Buras:1989xd,Buras:1991jm,Buras:1992tc,Ciuchini:1993ks,Ciuchini:1993fk}. Differences can though remain in presence of UV divergencies of the UV theory \cite{OlgosoRuiz:2024dzq}.

\section{Conclusion \label{sec:conclusion}}
In this paper we presented the two-loop contributions of $\order{g_s^2}$ of four-top operators to the running of the top quark mass and the strong coupling. Although these operators are not the only four-fermion operators contributing to the running of $m_t$ and $g_s$, we assume that their contributions could potentially be the most important ones as they are far less constrained than four fermion operators with light quarks. 

Since the RGEs are determined by the singularities associated to the Green's functions, we computed the divergent parts of the two-loop corrections to the top and the gluon propagator. The off-shell renormalization procedure requires the introduction of redundant operators, the choice of which can alter the final result. For this reason, we report in App.~\ref{app:Amplitude} a more general result that can be used for comparisons with different bases. 
We confirm the findings of Ref.~\cite{DiNoi:2023ygk} for what concerns the running of the strong coupling, presenting for the first time (at the best of our knowledge) the top-quark two-loop contribution to the running of the top mass. 

The computation has been performed by means of standard tools in the field of loop computations, namely using projectors and IBP identities to reduce the expression to a basis of scalar MIs. All the required integrals were available in literature. 

We treat our integrals with dimensional regularization, which is known to bring about several technicalities in the presence of chiral couplings. 
We perform our computation using the NDR scheme, whose application is well-defined for our purpose due to the small number of $\gamma$ matrices appearing in the expressions. 
By taking advantage of the relations first presented in Ref.~\cite{DiNoi:2025uan} we can convert our NDR results into the BMHV scheme. This last point is of particular interest since it is the only scheme which has been proved to be algebraically consistent in dimensional regularization, albeit introducing a plethora of practical complications. For this reason, obtaining the result in the BMHV scheme avoiding the challenges of an explicit computation can represent a significant shortcut applicable to a wide range of scenarios.
\section*{Note added}
We compared partial results in the BMHV scheme for some operators with Ref.~\cite{Nateropinprep} whose results we kindly obtained before publication. The results depends on several scheme choices, for instance, regarding the redundant operators and the symmetry restoring counterterms. These choices and the differences between the SMEFT and the LEFT did not allow for a complete cross check. We were though able to confirm the vanishing result for the running of $g_s$ and the $\order{m_b^2 m_t}$ contribution in Eq.~\eqref{eq:betamtBMHV}, which does not depend on these choices. 
\section*{Acknowledgments}
We would like to thank Anisha, Karim Elyaouti, Francesco Moretti, Luca Naterop, Marko Pesut, Maurice Sch\"ußler, Peter Stoffer and Eleni Vryonidou for useful discussions. 
We are particularly grateful to Giuseppe Ventura for the fruitful conversations regarding the two-loop renormalization program in the SMEFT framework and to 
Pablo Olgoso for insightful comments on the scheme choices in EFTs.
This work received funding by the INFN Iniziativa Specifica APINE and AMPLITUDES and by the University of
Padua under the 2023 STARS Grants@Unipd programme (Acronym and title of the project:
HiggsPairs – Precise Theoretical Predictions for Higgs pair production at the LHC). This work was also partially supported by the Italian MUR Departments of Excellence grant 2023-2027 “Quantum Frontiers”.

\appendix
\section{Two-point functions} \label{app:Amplitude}
We present here the intermediate result of the combination of the bare two-loop diagrams (see Fig.~\ref{fig:props}) and the one-loop insertion of the one-loop counterterms (see Fig.~\ref{fig:propsct}), restricting ourselves to the divergent parts only. 
The one-loop insertions of the off-shell counterterm in Eq.~\eqref{eq:ctPenguin} have been already subtracted, yielding a local divergence, but no choice for the other redundant operators has been made.

We have, defining $ L_2 \equiv  { g_s^2}/{(16 \pi^2)^2}$ for brevity, for the gluon propagator: 
\begin{equation}\label{eq:Pi}
\begin{aligned}
\frac{\Pi_{\mathrm{T}}(s)}{L_2} =\frac{1}{\Lambda^2}\Big( & \frac{4 s^2}{9\epsilon^2} (2 \coeff{QQ}{(1)}+6  \coeff{QQ}{(3)}+ \coeff{tt}{}+ \coeff{Qt}{(8)})  \\
 +&\frac{4 s^2}{9\epsilon} (2 \coeff{QQ}{(1)}+6  \coeff{QQ}{(3)}+ \coeff{tt}{}) \\ 
 -&  \frac{4 s \,m_t^2}{ \epsilon} \Big(\coeff{Qt}{(1)}-\frac{1}{6}\coeff{Qt}{(8)} \Big) \,
\Big).
\end{aligned}
\end{equation}
The fermion propagator reads
\begin{equation}
\begin{aligned}
\frac{\Sigma_R (s)}{ L_2}&=\frac{1}{\Lambda^2} \Big( \frac{4}{9\epsilon^2} (s-3 m_t^2 )(\coeff{Qt}{(8)}+2 \coeff{tt}{}) \nonumber \\
 +& \frac{4m_t^2}{9 \epsilon}(9\coeff{Qt}{(1)}-2\coeff{Qt}{(8)}-7 \coeff{tt}{})  \\ 
+& \frac{2 s}{9 \epsilon} (2\coeff{tt}{} - \coeff{Qt}{(8)}) 
\Big), \label{eq:SigmaR}  
\end{aligned}
\end{equation}
\vspace{-25 pt}
\begin{equation}
\begin{aligned}
\frac{\Sigma_L (s)}{L_2} =\frac{1}{\Lambda^2}& \Big( \frac{2}{9\epsilon^2} (s-3 m_t^2 )(\coeff{Qt}{(8)}+4 \coeff{QQ}{(1)}+12 \coeff{QQ}{(3)}) \nonumber \\
 -& \frac{m_t^2}{9 \epsilon}(7\coeff{Qt}{(8)} -36\coeff{Qt}{(1)} + 28 \coeff{QQ}{(1)} + 84 \coeff{QQ}{(3)})   \\ 
-& \frac{ s}{9 \epsilon} ( \coeff{Qt}{(8)} - 4 \coeff{QQ}{(1)}-12 \coeff{QQ}{(3)}) \Big),\\ 
\quad & \quad 
\label{eq:SigmaL} 
\end{aligned}
\end{equation}

\begin{equation}
\begin{aligned}
\frac{\Sigma_S (s)}{L_2} = \frac{1}{\Lambda^2}&\Big( \frac{2m_t^2}{9\epsilon^2}  (311\coeff{Qt}{(8)}+240\coeff{Qt}{(1)}  \nonumber \\ &-12 \coeff{QQ}{(1)}-36 \coeff{QQ}{(3)}-12\coeff{tt}{}) \nonumber \\
 +& \frac{4m_b^2}{\epsilon}(8 \coeff{QQ}{(3)}+\coeff{Qt}{(8)})   \\ 
+& \frac{4 m_t^2}{9 \epsilon} (38 \coeff{QQ}{(1)} + 42 \coeff{QQ}{(3)} + 38 \coeff{tt}{} \nonumber \\ &- 6 \coeff{Qt}{(1)}-65 \coeff{Qt}{(8)})  \nonumber\\ 
-& \frac{s}{3 \epsilon} (4 \coeff{QQ}{(1)}+12 \coeff{QQ}{(3)}+4 \coeff{tt}{}+12 \coeff{Qt}{(1)}+\coeff{Qt}{(8)})
\Big), \label{eq:SigmaS}  
\end{aligned}
\end{equation}

\vspace{25 pt}
\section{Results in the on-shell basis}
\label{app:OSBasis}
In the on-shell basis, redundant operators are removed by using the EOMs, yielding a result which does not depend on the precise definition of the reduntant operators themselves. We recall that, when employing this basis, the inclusion of all the operators which enter a given process with external on-shell states is required. 
The $p^2$ dependence in the two-point function is avoided by choosing $p^2 = m_t^2$. 

The RGEs for the top mass in the on-shell scheme read

\begin{equation} \label{eq:betamtNDROS}
\begin{split}
\beta_{m_t}^{\mathrm{NDR}} =  L_2 \frac{m_t}{\Lambda^2} \frac{16}{9} & \Big(
9m_b^2(8 \coeff{QQ}{(3)}+8 \coeff{Qt}{(8)}) \\ 
+& 2 m_t^2 (16 \coeff{QQ}{(1)} + 16 \coeff{tt}{} + 12 \coeff{QQ}{(3)} \\ 
& \quad \quad - 3  \coeff{Qt}{(1)}-34 \coeff{Qt}{(8)}) 
\Big),
\end{split}
\end{equation}
\begin{equation} \label{eq:betamtBMHVOS}
\begin{split}
\beta_{m_t}^{\mathrm{BMHV}} =  L_2 \frac{m_t}{\Lambda^2} \frac{16}{9} & \Big(
9m_b^2(8 \coeff{QQ}{(3)}+8 \coeff{Qt}{(8)}) \\ 
+& 3 m_t^2 (4 \coeff{QQ}{(1)} + 8 \coeff{tt}{}  \\ & \quad +6 \coeff{Qt}{(1)}-15 \coeff{Qt}{(8)}) 
\Big). 
\end{split}
\end{equation}
The RGE of $g_s$ does not change with respect to the off-shell case, namely it is given by Eq.~\eqref{eq:betagsNDR} in NDR and it vanishes in BMHV.
\vfill
\bibliography{bibliography}
\end{document}